\documentclass{aastex}
\usepackage{emulateapj5}
\usepackage{natbib}
\slugcomment{To appear in \apj}
\shorttitle{Circular polarization from synchrotron sources}
\shortauthors{M. Ruszkowski and M.C. Begelman}

\begin{document}
\title{Circular polarization from stochastic synchrotron sources}

\author{Mateusz Ruszkowski and Mitchell C. Begelman\altaffilmark{1}}
\affil{JILA, Campus Box 440, University of Colorado, Boulder CO 80309-0440}
\email{mr@jila.colorado.edu; mitch@quixote.colorado.edu}

\altaffiltext{1}{Also at Department of Astrophysical and Planetary
Sciences, University of Colorado}

\begin{abstract}
The transfer of polarized radiation in stochastic synchrotron
sources is explored by means of analytic treatment
and Monte Carlo simulations. We argue that 
the main mechanism responsible for the circular polarization
properties of compact synchrotron sources is likely to be 
Faraday conversion and
that, contrary to common expectation,
 a significant rate of Faraday rotation does not necessarily 
imply strong depolarization.
The long-term persistence of the sign of circular polarization, observed in
many sources,
is most likely due to a small net magnetic flux generated in the central
engine, carried 
along the jet axis and superimposed on a highly turbulent magnetic field.
We show that the mean levels of circular and linear polarizations depend on
the number of field reversals along the line of sight and that the gradient
in Faraday rotation across turbulent regions can lead to 
``correlation depolarization''.
Our model is potentially applicable to a wide range of synchrotron sources. 
In particular, we demonstrate how our model can naturally explain the
excess of circular over linear polarization in the Galactic Center and the
nearby spiral galaxy M81 and discuss its application to the quasar 3C 279, 
the intraday variable blazar PKS 1519-273 and the X-ray binary SS 433.
\end{abstract}


\section{Introduction}
Polarization has proven to be an important tool in AGN research. 
In principle, linear and particularly circular polarization 
observations of synchrotron radiation may permit 
measurements of various properties of jets such as: magnetic field strength
and topology, the net magnetic flux carried by jets (and hence
generated in the central engine), the energy spectrum of radiating
particles,  
and the jet composition (i.e., whether jets are mainly composed of 
$e^{+}-e^{-}$ pairs or electron-proton plasma).
The renewed interest in polarization of compact radio sources stems from
two recent developments. 
First, \citet{bow99} detected circular polarization using the Very Large 
Array (VLA) in
the best supermassive black hole candidate, the Galactic center (Sgr
A$^{*}$). This discovery was quickly confirmed by \citet{sau99} using 
the Australia Telescope Compact Array (ATCA).
Circular polarization was also detected in  
the celebrated X-ray binary system SS 433 \citep{fen00}.
Moreover, the Very Long Baseline Array (VLBA) has
now detected circular polarization in as many as 20 AGN \citep{war98,hom99}.
Second, it is now possible to measure circular
polarization with unprecedented accuracy of 0.01\% using 
the ATCA \citep{ray00}.
This dramatic improvement in the observational status of polarization measurements
 has also brought new questions.
For example, there is now growing observational evidence that the sign of
circular polarization is persistent over decades \citep{kom84,hom99}, 
which indicates that it
is a fundamental property of jets. Another problem, which has not been
satisfactorily explained as yet, is how to reconcile the high level of circular 
polarization with the lower value of linear polarization in Sgr A$^{*}$
\citep{bow99} and M81$^{*}$ \citep{bru01}.
Indeed, there is not even a general consensus on the mechanism
responsible for the circular polarization properties of jets \citep{war98}. \\
\indent
In this paper we attempt to solve some of the theoretical puzzles.
The paper is organized as follows. In the next section we summarize
the most important observational facts. In Section 3 we briefly discuss 
mechanisms for producing circular polarization and argue that the most
likely process is ``Faraday conversion''. Section 4 presents our model for
polarization and in the subsequent sections we compare 
Monte Carlo simulations with
analytic results and discuss general implications for observations 
as well as specific observational cases.
We summarize our conclusions in Section 7.

\section{Observational trends}
Compact radio sources typically show a linear polarization (LP) of a 
few percent of the total intensity \citep{jon85}. This is much less than the 
theoretical maximum for synchrotron sources, which can approach $70\%$ in 
homogeneous sources with unidirectional magnetic field. Therefore, magnetic
fields in radio sources are believed to be highly inhomogeneous, although the
nonvanishing linear polarization is in itself an indirect indication of
a certain degree of ordering of the field. 
Although the precise topology of the magnetic
field in jets is not known, there are other compelling reasons, both theoretical
and observational, to believe that magnetic fields are indeed partially
ordered. On the observational side, these conclusions are based on
measurements of the orientation of linear polarization ($\mathbf E$
vector), which reveal coherent structures across jet images. 
This indicates
that magnetic fields, which are predominantly perpendicular to the electric
vectors, are also preferentially aligned, although in different sources or
in different parts of the jet
the magnetic fields can be mainly orthogonal \citep{war98} 
or parallel \citep{jon85,rus85} to the
projected jet orientation. From the theoretical
point of view, ordered jet magnetic field is expected when 
shocks compress an initially random field \citep{lai80,lai81,mer85,hug89,war94} 
($\mathbf B$ perpendicular to the jet axis) or when such initial
fields are sheared to lie in a plane \citep{lai80,lai81,beg84} 
($\mathbf B$ parallel to the jet axis). \\
\indent
Circular polarization (CP) is a common feature of quasars and blazars
\citep{ray00,hom01}, is usually characterized by an approximately flat
spectrum, and is generated near synchrotron self-absorbed jet cores
\citep{hom99}. 
CP is detected in about 30\%-50\% of these objects. 
Measured degrees of CP are generally lower than the levels of linear polarization and
usually range 
between 0.1 and 0.5\% \citep{hom99,hom01}. As reported by \citet{mar00} for
the intraday variable source PKS 1519-273, the CP of variable components of
intensity can be much higher than the overall circular
polarization levels. Observations of proper motion of CP-producing
regions in the quasar 3C 273 \citep{hom99} suggest that circular polarization is
intrinsic to the source, as opposed to being due to foreground effects. 
Most
importantly, comparisons of CP measurements made within the last 30 years
\citep{wei83,kom84,hom99} with the most recent observations reveal that, despite
CP variability, its sign is a persistent feature of AGN, which must
therefore be
related to a small net unidirectional component of magnetic field in jets.

\section{Mechanisms for producing circular polarization}
The most obvious candidate for explaining circular 
polarization of compact radio sources is intrinsic emission \citep{leg68}. 
Although intrinsic CP is roughly $\pi_{c, \rm int}\sim\gamma^{-1}$
where $\gamma$ is the Lorentz factor of radiating electrons, in a
realistic source it will most likely be strongly 
suppressed by the tangled magnetic field and
possibly the emissivity from $e^{+}-e^{-}$ pairs, which do not contribute
CP. Specifically, 
$\pi_{c, \rm int}\sim\gamma^{-1}(B_{u}/B_{\rm rms})f_{\rm pair}\ll 1\%$, where 
$B_{u}$ and $B_{\rm rms}$ are the unidirectional component of the magnetic
field projected onto the line-of-sight and 
the fluctuating component of the field, respectively, and  
$f_{\rm pair}\equiv(n^{-}-n^{+})/(n^{-}+n^{+})\le 1$. Therefore, intrinsic CP 
appears to be inadequate to explain the observed polarization.
Other mechanisms have also been proposed, among which the most popular ones
are coherent radiation processes \citep{ben00}, 
scintillation \citep{marm00} and Faraday conversion \citep{pac75,jon77a,jon88,war98}.
The first of these mechanisms produces polarization in a narrow frequency
range which now seems to be ruled out by multiband observations. The recently
proposed scintillation mechanism, in which circular
polarization is stochastically produced by a birefringent screen located
between the jet and the observer, fails to explain the persistent sign of circular
polarization as the time-averaged CP signal is predicted to vanish. 
The last mechanism --- Faraday conversion --- seems to be the most
promising one and in the next subsection we discuss it in more detail. 

\subsection{Faraday rotation and conversion}
The polarization of radiation changes as it propagates through any
medium in which modes are characterized by different plasma speeds.
In the case of cold plasma the modes are circularly
polarized. The
left and right circular modes have different phase velocities
and therefore the linear polarization vector of the propagating
radiation rotates. This effect is called Faraday rotation and it is often used to
estimate magnetic field strength in the interstellar medium and to estimate
pulsar distances. Note that Faraday rotation does not alter the degree of
circular polarization, since any circular polarization can be decomposed into
two independent linearly polarized waves.
Faraday rotation is a specific example of a more general phenomenon
called birefringence. In a medium whose natural modes are linearly or
elliptically polarized, such as a plasma of relativistic particles,
birefringence leads to the partial cyclic conversion between linearly and
circularly polarized radiation as the phase relationships between the modes
along the ray change with position. 
This effect is best visualized by means of the 
Poincar\'{e} sphere \citep{mel91,ken98}. An arbitrary elliptical polarization
can be represented by a 
vector $\mathbf P$ with its tip lying on the Poincar\'{e}
sphere and characterized by Cartesian coordinates
$(Q,U,V)/I$, where $Q,U,V$ and $I$ are the Stokes parameters (see Fig. 1). 
Thus, the north
and south poles correspond to right and left circular
polarizations and points on the equator to linear polarization.
Different azimuthal positions on the sphere correspond to different
orientations of the polarization ellipses.
The polarization of natural modes of the medium is represented by a diagonal
axis, whose polar angle measured from the vertical axis 
depends on whether the medium is dominated by
cold $(0^{\rm o})$ or highly relativistic particles $(90^{\rm o})$. 
As the radiation passes through the
medium, birefringence causes the tip of the polarization vector to rotate
at a constant latitude around the axis of the natural plasma modes. 
In this picture, Faraday rotation
corresponds to the case where the natural modes axis is vertical and
the polarization vector $\mathbf P$ rotates around it. Note that, even if
radiation initially has no circular polarization (i.e., $\mathbf P$ lies in the
equatorial plane) and then encounters a medium in which the normal modes are
elliptical, it will develop an elliptically polarized component. 
An interesting property of a relativistic
birefringent plasma is that it can generate circular polarization even if 
it is composed almost entirely of electron-positron pairs. At first this may seem
paradoxical, as one would expect electrons and positrons to contribute
to CP with opposite signs. However, despite the fact that the intrinsic
CP in such a case is indeed close to zero, some additional 
CP can be produced by conversion of linear polarization
\citep{saz69,noe78}. In terms of the Poincar\'{e} sphere, this situation
corresponds to the normal modes axis pointing close to but not exactly in
the equatorial plane. Therefore some conversion of intrinsic linear
polarization may occur provided that there is some imbalance in the number
of electrons and positrons.

\subsubsection{Strong rotativity limit}
Strong departures from
mode circularity occur only when radiation propagates within a small angle
$\sim\nu_{L}/\nu$ of the direction perpendicular to the magnetic
field, where $\nu_{L}=eB/2\pi m_{e}c$.
Therefore radiative transfer 
is often performed in the quasi-longitudinal (QL) approximation. 
If the normal modes are highly elliptical then the opposite,
quasi-transverse (QT), limit applies \citep{gin61}.
In a typical observational situation it is usually assumed that Faraday
rotation within the source cannot be too large, as this will lead to the 
suppression of linear
polarization. However, this constraint does not prevent rotativity from
achieving large values locally as long as the mean rotativity, i.e., averaged over
all directions of magnetic field along the line of sight, 
is indeed relatively small. Such a situation may happen in a turbulent plasma. 
Some effects of turbulence on polarization were discussed by 
\citet{jon88} who neglected a uniform magnetic field component and by
\citet{war98}, who presented results for the case of a small synchrotron depth.
Technically, the strong rotativity regime is equivalent to the QL limit
and in this paper we build our model on this approximation.

\section{Model for polarization}
We consider a highly tangled magnetic field with a very small mean
component which is required to determine the sign of circular polarization.
From a theoretical view-point, we would expect some net poloidal magnetic
field, either originating from the central black hole or from the accretion
disk, to be aligned preferentially along the jet axis. 
Specifically, from equipartition and flux freezing
arguments applied to a conical jet 
\citep{bla79} we get $\langle B^{2}_{\|}\rangle^{1/2}\sim\langle
B^{2}_{\bot}\rangle^{1/2}\sim B_{\rm rms}\propto r^{-1}$ where 
$r$ is the distance along the synchrotron emitting source and 
the symbols $\|$
and $\bot$ refer to magnetic fields parallel and perpendicular to the jet axis,
respectively. From the flux-freezing
argument applied to the small parallel bias in the magnetic field we obtain 
$\langle B_{\bot}\rangle\sim 0$ and $\langle B_{\|}\rangle\propto
r^{-2}\propto\delta B_{\rm rms}$, where $\delta\equiv B_{u}/B_{\rm rms}\ll
1$ 
is the ratio of the uniform and fluctuating components of the magnetic field.

\subsection{Mean Stokes parameters in the presence of field reversals}
We solve the radiative transfer of polarized radiation in a turbulent plasma 
by adopting transfer equations for a piecewise homogeneous 
medium with a weakly anisotropic dielectric tensor \citep{saz69,jon77a}. 
Details of the transfer equations 
are given in the Appendix. We assume that the mean rotativity per unit
synchrotron optical depth
$\langle\zeta_{v}^{*}\rangle\equiv\delta\zeta$ and that
$\langle\sin2\phi\rangle =
0$ and $\langle\cos2\phi\rangle =2p-1$, where $0\leq p\leq 1$ is a
parameter describing the polarization direction and degree of order in the
field. We also assume that
circular absorptivity $\zeta_{v}$ and circular emissivity
$\epsilon_{v}$ are both negligible. 

\subsubsection{Large synchrotron depth limit}
Averaging the transfer equations over orientations of the magnetic field,
we obtain the following asymptotic expressions for large synchrotron optical 
depth:\\
\begin{eqnarray}
\overline{I}+(2p-1)\zeta_{q}\overline{Q} & = & J\\
\overline{Q}+\langle\zeta_{v}^{*}U\rangle +(2p-1)\zeta_{q}\overline{I} & = & (2p-1)\epsilon_{q}J\\
\overline{U}-\langle\zeta_{v}^{*}Q\rangle +(2p-1)\overline{\zeta}_{q}^{*}\overline{V} & = & 0\\
\overline{V}-(2p-1)\overline{\zeta}_{q}^{*}\overline{U} & = & 0 \; .
\end{eqnarray}
Note that $\zeta_{v}^{*}$ is not statistically independent from $U$
and $Q$ due to the gradient in Faraday rotation across each cell. The correlation 
in eq. (2) then reads:\\
\begin{equation}
\langle\zeta_{v}^{*}U\rangle=\delta\zeta\overline
U+\langle\widetilde{\zeta}_{v}^{*}\widetilde{U}\rangle,
\label{eq5}
\end{equation}
where $\widetilde{\zeta}_{v}^{*}$ and $\widetilde{U}$ denotes the
fluctuating part of $\zeta_{v}^{*}$ 
and $U$. An analogous relation holds for $Q$ in eq. (3). 
We neglect the term
 $\langle\widetilde{\zeta}_{q}^{*}\widetilde{U}\rangle$ in eq. (4) as 
convertibility is a much weaker function of plasma
parameters than rotativity.
It will be shown in Section 4.1.3 that the
correlations $\langle\widetilde{\zeta}_{v}^{*}\widetilde{U}\rangle$ and
$\langle\widetilde{\zeta}_{v}^{*}\widetilde{Q}\rangle$
tend to zero as the number of field reversals along the line of
sight increases and that\\
\begin{equation}
-\langle\widetilde{\zeta}_{v}^{*}\widetilde{Q}\rangle/\overline{U}=
\langle\widetilde{\zeta}_{v}^{*}\widetilde{U}\rangle/\overline{Q}\equiv\xi
\; .
\end{equation}
This implies that the mean levels of
circular and linear polarizations will also depend on the
number of the field reversals along the line of sight.
Setting $\overline{\pi}_{q}=\overline{Q}/\overline{I}$,  
$\overline{\pi}_{u}=\overline{U}/\overline{I}$ and 
$\overline{\pi}_{v}=\overline{V}/\overline{I}$, we get mean normalized 
Stokes parameters from equations (1)--(4):\\ 
\begin{eqnarray}
\overline{\pi}_{q}&=&\frac{(2p-1)[1-\xi
+(2p-1)^{2}\overline{\zeta}_{q}^{*2}](\epsilon_{q}-\zeta_{q})}{\cal D}\\
\overline{\pi}_{u}&=&\frac{(2p-1)\delta\zeta(\epsilon_{q}-\zeta_{q})}{\cal D}\\
\overline{\pi}_{v}&=&\frac{(2p-1)^{2}\overline{\zeta}_{q}^{*2}\delta\zeta(\epsilon_{q}-\zeta_{q})}{\cal D},
\end{eqnarray}
where 
\begin{eqnarray}
{\cal D}&\equiv&1+2\xi+(2p-1)^{2}(\zeta_{q}^{*2}-\zeta_{q}\epsilon_{q})+
\delta^{2}\zeta^{2}+\nonumber\\
 & &[\xi-(2p-1)^{2}\zeta_{q}\epsilon_{q}]
[\xi+(2p-1)^{2}\overline{\zeta}_{q}^{*2}] \; ,
\end{eqnarray}
Note that\\
\begin{equation}
\frac{\overline{\pi}_{v}}{\overline{\pi}_{u}}=(2p-1)\overline{\zeta}^{*}_{q}
\end{equation}
\begin{equation}
\frac{\overline{\pi}_{v}}{\overline{\pi}_{q}}=
\frac{(2p-1)\overline{\zeta}^{*}_{q}\delta\zeta}{1+\xi+
(2p-1)^{2}\overline{\zeta}^{*2}_{q}} \; .
\end{equation}
Therefore, circular polarization $\overline{\pi}_{c}=-\overline{\pi}_{v}$
can dominate over linear polarization 
$\overline{\pi}_{l}=(\overline{\pi}^{2}_{q}+\overline{\pi}^{2}_{u})^{1/2}$
if $\delta\zeta\ga (2p-1)\zeta^{*}_{q}>1$, where the
last inequality holds for a large number of field reversals along the line
of sight.
\subsubsection{Small synchrotron depth limit}
In the limit of small synchrotron depth and for $\delta\zeta\ga 1$, 
we approximately have:
\begin{eqnarray}
\frac{d\overline{I}}{d\tau}&=&J\\
\frac{d\overline{Q}}{d\tau}+\delta\zeta\overline{U}+\langle\widetilde{\zeta}_{v}^{*}\widetilde{U}\rangle&=&(2p-1)\epsilon_{q}J\\
\frac{d\overline{U}}{d\tau}-\delta\zeta\overline{Q}-\langle\widetilde{\zeta}_{v}^{*}\widetilde{Q}\rangle&=&0\\
\frac{d\overline{V}}{d\tau}-(2p-1)\overline{\zeta}_{q}^{*}\overline{U}&=&0
\; .
\end{eqnarray}
Introducing $X\equiv \overline{Q}+i\overline{U}$, where $i=\sqrt{-1}$, we
obtain:
\begin{equation}
\frac{dX}{d\tau}+(\xi -i\delta\zeta)X=(2p-1)\epsilon_{q}J
\end{equation}
Solving eq. (16) and using eq. (17) we get:
\begin{eqnarray}
\overline{\pi}_{l}&=&(2p-1)\epsilon_{q}\left [
\frac{1-2e^{-\tau_{\xi}}\cos\tau_{r}+e^{-2\tau_{\xi}}}
{\tau_{\xi}^{2}+\tau_{r}^{2}}\right ]^{1/2}\\
\overline{\pi}_{v}&=&(2p-1)^{2}\epsilon_{q}\frac{\tau_{c}}{\tau_{\xi}^{2}+\tau_{r}^{2}}\times
\nonumber \\
&&
\left
[\tau_{r}+\frac{[2\tau_{r}\tau_{\xi}\cos\tau_{r}+(\tau_{\xi}^{2}-\tau_{r}^{2})\sin\tau_{r}]e^{-\tau_{\xi}}-2\tau_{r}\tau_{\xi}}{\tau_{\xi}^{2}+\tau_{r}^{2}}\right
],
\end{eqnarray}
where $\tau_{r}=\delta\zeta\tau$, $\tau_{c}=\zeta_{q}^{*}\tau$ and 
$\tau_{\xi}=\xi\tau$ are the rotation, conversion and ``correlation'' depths,
respectively. 
As in the synchrotron thick case, circular polarization can exceed linear
polarization. For example, when $\tau_{\xi}$ is negligible, the CP/LP ratio
exceeds unity if:\\
\begin{equation}
(2p-1)\tau_{c}\ga 2\left|\sin\left(\frac{\tau_{r}}{2}\right)\right|.
\end{equation}
As equations (6)--(9) for the synchrotron thick case and equations
(18) and (19) for the synchrotron thin case clearly demonstrate, 
correlations induced by the gradient in Faraday rotation across turbulent
regions have a depolarizing effect on linear and circular polarizations.

\subsubsection{Effect of statistical fluctuations on the mean Stokes 
parameters}
Having obtained averaged quantities, we now proceed to calculate the
effect of statistical fluctuations on the mean Stokes parameters. 
We introduce the
following notation for the mean and fluctuating parts of the Stokes
parameters: $S=\overline{S}+\widetilde{S}$, where 
$\langle \widetilde{S}_{i}\rangle =0$; and
for the angular distribution of the projected magnetic field: 
$\sin 2\phi\equiv a$, $\cos 2\phi =(2p-1)+b$,
where $\langle a\rangle =\langle b\rangle =\langle
ab\rangle =0$. We write the fluctuating rotativity as 
$\zeta_{v}^{*}\equiv\delta\zeta +\widetilde{\zeta}$, where
$\langle\widetilde{\zeta}\rangle =0$ 
but $\widetilde{\zeta}\sim{\cal O}(\zeta)$. We constrain ourselves to the case
dominated by Faraday rotation, i.e., we have $|\zeta Q|\gg
|a\zeta_{q}I|$, 
etc. In such a case, the lowest order terms do not depend on the
fluctuations in the orientation of the projected magnetic field $\phi$.
Thus, retaining only leading terms, 
we obtain the fluctuating part of the transfer equations for 
$\widetilde{Q}$ and $\widetilde{U}$:

\begin{equation}
\frac{d\widetilde{Q}}{d\tau}+\widetilde{\zeta}(\overline{U}+\widetilde{U})=0
\end{equation}
\begin{equation}
\frac{d\widetilde{U}}{d\tau}-\widetilde{\zeta}(\overline{Q}+\widetilde{Q})=0
\; .
\end{equation}
Note that when the fluctuations in $Q$ and $U$ are not dominated by 
rotativity terms, they are determined by variations  
in the orientation of the magnetic field. 
From equations (21) and (22) we can obtain 
corrections to the mean Stokes parameters due
to the gradient in Faraday rotation across each cell:\\ 
\begin{equation}
\langle\widetilde\zeta_{v}^{*}\widetilde{Q}\rangle=
-\left\langle\frac{\widetilde{\Delta\tau}}{2}\widetilde{\zeta}^{2}(\overline{U}+\widetilde{U})\right\rangle=-\frac{1}{2}\overline{U}\langle\widetilde{\zeta}^{2}\widetilde{\Delta\tau}\rangle
\end{equation}
\begin{equation}
\langle\widetilde\zeta_{v}^{*}\widetilde{U}\rangle=
\frac{1}{2}\overline{Q}\langle\widetilde{\zeta}^{2}\widetilde{\Delta\tau}\rangle,
\end{equation}
where $\widetilde{\Delta\tau}=\Delta\tau (\sin\theta)^{\alpha +3/2}$,
$\theta$ is the angle between the line of sight and the direction of the
magnetic field, $\Delta\tau =\tau_{\rm o}/N$, $\tau_{\rm o}$ is
the maximum synchrotron depth and $N$ is the number of turbulent zones
along the line of sight within this depth.
Although these terms play an important role in the expressions for the mean
Stokes parameters (equations [6]--[10] and [18]--[19]), they
introduce only higher order corrections 
$\sim {\cal O}(\zeta^{2}\Delta\tau/2)$ to the fluctuation
equations and therefore should not be taken into account in equations (21) and
(22) for the treatment
to be self-consistent (recall that we assume that $\zeta\Delta\tau <1$). Note
also that both $\langle\widetilde\zeta_{v}^{*}\widetilde{Q}\rangle$ and 
$\langle\widetilde\zeta_{v}^{*}\widetilde{U}\rangle$ are proportional
to $\tau_{\rm o}N^{-1}$ and therefore asymptotically tend to zero as the
number of field reversals along the line of sight increases.\\
\indent
We note that the presence of inhomogeneities in the magnetic field can, under certain
circumstances, introduce additional complications into the radiative transfer
due to the tracking and coupling of plasma modes. 
As the propagating wave goes from the QL to
the QT regime (i.e. $\mathbf B$ almost perpendicular to the line of
sight) and then back to the QL regime, it can adiabatically adjust itself to the shifting
nature of the eigenmodes provided that $\zeta_{ov}\Delta\tau\gg 1$, where 
$\zeta_{ov}\equiv\zeta_{\alpha}^{*v}\gamma^{2}\ln\gamma_{i}\;/\gamma_{i}^{3}$
\citep{bjo90,tho94,fuk98}, i.e., provided that the gradient in the magnetic
field is sufficiently small. In the opposite limit, the propagating wave does not
`notice' any inhomogeneities. In such a case, a circularly
polarized wave will preserve its helicity as it crosses the QT region. 
Therefore, our assumption that Stokes $I$, $Q$, $U$ and, in particular, $V$,
are continuous at sharp boundaries between the turbulent cells, is
consistent with our initial assumption that $\zeta_{ov}\Delta\tau <1$. 
A spatially varying magnetic field can also introduce mode coupling when
the coupling constant ${\cal L}\sim
\zeta_{oq}/(\zeta_{ov}^{2}\Delta\tau)\sim 1$,
where $\zeta_{oq}\equiv 2\zeta_{\alpha}^{*q}\ln (\gamma/\gamma_{i})$
\citep{jon77b} even in the QL region (i.e., $\mathbf B$ not in a
direction almost perpendicular to the line of sight). 
When ${\cal L}\ll 1$
and propagation occurs in the QL regime, coupling effects are unimportant. In
the opposite limit when ${\cal L}\gg 1$, and when there is
no uniform component of the magnetic
field, radiation propagates as in a vacuum. Although we present no formal
proof, we argue that in the latter limit radiation will
be unaffected by the fluctuating component of the magnetic field and will
be sensitive only to the mean bias component of the total magnetic field.
In a real situation, the precise value of ${\cal L}$ will depend on the
details of magnetohydrodynamical turbulence.
However, since our calculations were performed assuming a piece-wise
homogeneous medium, technically our results are exact. This is consistent
with claims made by \citet{jon88}, who also considered the piece-wise
homogeneous case and performed calculations using ``standard'' transfer
equations and the wave equation. He 
obtained practically identical results from the two methods even
though some of his results formally violated the non-coupling criterion.
In a real situation we would expect some gradient of magnetic field across
each turbulent cell but, again, in general the results will depend on the detailed treatment of
the MHD turbulence which is beyond the scope of this work.

\section{Results and implications}
We now consider a range of specific examples aimed at demonstrating
predictions of our model and the consistency of our formulae with the results
of Monte Carlo simulations.
As mentioned in Section 3.1.1, we focus on cases where Faraday rotation per
unit synchrotron optical depth $\zeta_{v}^{*}$ is large.
Specifically, we assume that the 
typical mean Lorentz factor of radiating electrons $\gamma\sim 10^{2}$ and that
the electron energy distribution function has a power-law form
$n(\gamma)\propto\gamma^{-(2\alpha +1)}$, where $\alpha$ is the spectral index
of optically thin synchrotron emission. We use $\alpha=0.5$ 
and assume that the electron distribution is cut-off
below $\gamma_{i}\sim$ a few. For example, for the maximum brightness
temperature $T_{b}\sim 10^{11}K$ \citep{rea94} we have $\gamma\sim 3kT_{b}/m_{e}c^{2}\sim
50$, which corresponds to mean rotation and conversion per unit
synchrotron optical depth of order $\sim \delta\zeta_{v}^{*}\sim 
3\times 10^{3}\delta\ln\gamma_{i}/\gamma_{i}^{3}$ and 
$\zeta_{q}^{*}\sim -\ln(\gamma/\gamma_{i})$, respectively, for
$\nu\sim\gamma^{2}eB/2\pi m_{e}c$ 
\footnote{We further comment on the choice of parameters and consider more
general situations while discussing specific observational cases in Sections
6.1 and 6.2}.
In order to facilitate comparison of
analytical and numerical results we use appropriate 
synchrotron depth-weighted solid angle
averages of coefficients of rotativity $\zeta_{v}^{*}$ and 
convertibility $\zeta_{q}^{*}$ which include the effect of
inclination of the uniform component of magnetic field to the line of
sight:
\begin{equation}
\overline{\zeta}_{q,v}^{*}=\frac{\langle\zeta_{q,v}^{*}
(B\sin\theta)^{\alpha +3/2}\rangle_{\Omega}}
{\langle (B\sin\theta)^{\alpha +3/2}\rangle_{\Omega}}.
\end{equation}
Retaining only the leading terms, we get
\begin{equation}
\overline{\zeta}_{q}^{*}=-2\zeta_{\alpha}^{*q}\ln\left(\frac{\gamma}{\gamma_{i}}\right)+\frac{3}{8}\pi
\left(\frac{1}{2}-\ln 2\right)\zeta_{\alpha}^{*q}
\end{equation}
\begin{equation}
\overline{\zeta}_{v}^{*}=\frac{3}{2}\zeta_{\alpha}^{*v}\gamma^{2}\frac{\ln\gamma_{i}}{\gamma_{i}^{3}}\delta\cos\theta_{u},
\end{equation}
where $\theta_{u}$ is the angle between the line of sight and the direction
of the uniform magnetic field component. Analogously, for the correlation
terms (equations [23] and [24]), we use
\begin{equation}
-\frac{\langle\widetilde{\zeta}_{v}^{*}\widetilde{Q}\rangle}{\overline{U}}=
\frac{\langle\widetilde{\zeta}_{v}^{*}\widetilde{U}\rangle}{\overline{Q}}=
\frac{\Delta\tau}{4}\left(\zeta_{\alpha}^{*v}\gamma^{2}\frac{\ln\gamma_{i}}{\gamma_{i}^{3}}\right)^{2}\equiv\xi
.
\end{equation}

\subsection{Mean linear and circular polarizations}
Fig. 2 presents mean linear (upper panel) and circular 
polarizations for the case of radiation transfer 
through a high synchrotron depth. 
The uppermost lines on each panel show analytic results for ``saturated'' polarizations corresponding to 
$\xi\la 1$ for $p=1$ (i.e., no dispersion of the
projected magnetic field on the sky), and filled squares connected by
dashed lines denote results of numerical simulations. Small systematic
differences on Figures 2 and 3 between the analytical curves and the
numerical ones at higher values of $\delta$ are due to the fact that the
analytical formulae do not include higher order $\delta$-terms. 
The general behavior of these curves can be
understood in simple terms. Linear polarization gradually decreases with
$\delta$ (and mean rotativity $\overline{\zeta_{v}^{*}}$) 
as a result of the increasing strength of Faraday
depolarization. Note that, contrary to common expectation, the mean linear
polarization does not vanish as the result of a highly inhomogeneous magnetic field 
even though the magnitude
of rotation per unit synchrotron optical depth is very large. As expected, circular
polarization initially becomes stronger with an increasing component of the
uniform magnetic field parallel to the line-of-sight. However, as 
Faraday depolarization gradually eliminates linear polarization (both
Stokes $Q$ and $U$ are affected by this process), there is a 
reduced amount of $U$ available for conversion into circular polarization
(Stokes $V$). Thus circular polarization has an extremum. \\
\indent
Other curves on
Fig. 2 represent the case of ``unsaturated'' polarization when $\xi\ga 1$. 
In such situations, the levels of both linear and circular
polarizations systematically decline as $N$ decreases. Note that the number
of turbulent zones along the line of sight cannot be arbitrarily low, as 
this would reduce the mean polarization levels to
very low values due to the increasing influence of the
correlations between rotativity and Stokes $Q$ and $U$ parameters.
Apart from affecting the mean polarization levels,
increasing the number of field reversals along the line of sight also 
reduces the polarization fluctuations. Due to the finite
size of the telescope beam, the fluctuations are further suppressed, however
the beam averaging process does not affect the mean level of circular 
polarization.
It also does not influence the mean level of linear polarization, provided
that LP fluctuations are not very large prior to the beam averaging.
Assuming that the number of turbulent zones in each of
the perpendicular directions across the telescope beam is very
roughly comparable to $N$, we conclude that fluctuations in
the mean levels of CP and LP due to the stochastic nature of the plasma 
should be relatively small. 
Indeed, even though our numerical results for the 
mean circular and linear 
polarizations were obtained for a number of ``pencil beams'' much smaller
than $N^{2}$, none of our results exhibit strong fluctuations.
Thus, by demanding that there be enough turbulent cells along the line of
sight to avoid correlation depolarization, we also guarantee that
the polarization variability should be dominated
by changes in the mean plasma parameters (e.g., the magnitude of the 
uniform component of the magnetic field, dispersion of the projected random
magnetic field on the sky, the synchrotron depth, etc.), rather than by 
statistical fluctuations in, e.g., the
orientation of the local turbulent magnetic field.
This also assures that the sign of circular polarization should be
a persistent feature --- variations of the mean plasma parameters are 
not likely to change the sense of circular polarization,
provided that the synchrotron optical depth does not change dramatically 
from low to very high values (see below) and the orientation of the mean field
remains the same.\\
\indent
Fig. 3 illustrates the effect of relaxing the assumption of 
that the the projected magnetic field has zero dispersion on the sky.
As the projected orientation of the magnetic field becomes
increasingly chaotic $(p\rightarrow 0.5)$, the magnitudes of 
linear and circular polarizations gradually decline.\\
\indent
As both Figures 2 and 3
indicate, circular polarization can exceed linear polarization at higher
values of the bias parameter $\delta$. 
The excess of CP over LP requires significant Faraday
depolarization of linear polarization. Nevertheless, 
even a small amount of $U$
can then be effectively converted to circular polarization, leading to CP/LP
ratios in excess of unity. It is even conceivable to have a situation in which
linear polarization falls below the detection threshold whereas circular
polarization is still easily observable.
Note that significant Faraday depolarization does not require a large
bias $\delta$ (i.e., exceeding unity)
because of the large values of rotativity. This is consistent with
observations which do not reveal any dominant large scale unidirectional 
fields. Such fields would lead to very strong linear polarization and 
are also unexpected on theoretical grounds.\\
\indent
In real sources the effective synchrotron optical depth will depend 
not only on the emission properties of the plasma 
along the line of sight but also, among other factors, 
on the solid angle subtended by the emitting region. 
In a realistic situation, the telescope will integrate over a finite-size beam with
different synchrotron depths along different lines of sight.
Thus, the emission from the synchrotron self-absorbed core of a jet
will be weighted by a smaller solid angle than the 
emission from the more extended regions
which have lower synchrotron depths and, therefore, 
lower surface brightness. Detailed calculations of 
the polarization properties of specific jet models are beyond the scope of this
paper and will be presented in a forthcoming publication (Ruszkowski and
Begelman, in preparation).
Fig. 4 illustrates the effect of varying the total synchrotron depth 
$\tau_{\rm o}$ for `saturated' polarization ($\xi\la 1$). 
For large $\tau_{\rm o}$
the results are similar to the ones presented on Fig. 2 and 3. In the
opposite extreme, i.e., when $\tau_{\rm o}$ tends to zero, we have 
qualitatively similar behavior with the main difference being that the helicity
of circular polarization is reversed. 
As in the high synchrotron depth case, the general trends in the
polarization behavior can be understood in simple terms. The gradual
decline of linear polarization with $\delta$ is just a result of Faraday
depolarization. 
As radiation propagates through the plasma, Stokes $U$ is 
generated from Stokes $Q$ due to Faraday rotation. Therefore, 
circular polarization, which is produced from Stokes $U$ by Faraday 
conversion, initially increases. However, for large $\delta$, Faraday
depolarization reduces Stokes $Q$ and $U$ and thus leads to the suppression
of circular polarization. 
The oscillations in linear and circular polarizations are
due to cyclic rotation of Stokes $Q$ into $U$ followed by
conversion to $V$. Note that, for a narrow range of 
intermediate values of synchrotron
depth, the behavior of CP resembles that of low $\tau_{\rm o}$
when $\delta$ (and the mean rotativity) is small, and that of high 
$\tau_{\rm o}$ when $\delta$ is larger.\\
\indent
The effects of ``correlation depolarization'' for synchrotron
depths around unity are showed in Fig. 5. This figure also demonstrates our
analytic results for small synchrotron depths (dashed lines).
Clearly, our analytic solutions for both very large and small
synchrotron depths adequately describe the numerical results which smoothly
join the two regimes.

\section{Application to specific sources}
\subsection{The case of quasar 3C279}
\citet{war98} reported the discovery of circular and linear polarization in
3C 279 and attributed CP to internal Faraday conversion. 
Typical fractional linear and circular polarizations in 3C 279 
are of order $\sim 10\%$ and $\la 1\%$, respectively.
\citet{war98} concluded that
if the jet is composed of normal plasma, then the low-energy cut-off of the
energy distribution of relativistic electrons must be as high as 
$\gamma_{i}\sim 100$ in order to avoid Faraday depolarization and 
overproduction of the jet kinetic power. They considered synchrotron thin
models with a bias in magnetic field and 
a small number of field reversals along the line of sight but
neglected the dispersion of the projected magnetic field on the sky.
They were unable to fit their polarization models to the observational data
for $\gamma_{i}\ga 20$ and thus claimed that the jet must be
pair-dominated. 
However, the above observational constraints on CP and LP and the 
jet energetics can be satisfied for a variety of microscopic plasma
parameters. This is due to the fact that 
different ``microscopic'' parameters, such as $\gamma_{i}$, the ratio of the cold to
relativistic electron number densities, or the positron fraction, can lead
to similar ``macroscopic'' parameters such as convertibility and rotativity.
In order to illustrate this, we consider two radically different examples 
and show that both cases can lead to the same CP and LP.
\subsubsection{Electron-proton jet}
In this example, plasma is composed exclusively of a mixture of protons and electrons 
with both relativistic and cold populations being present.
For instance, for a low-energy cut-off
$\gamma_{i}\sim 30$ and an electron number density-weighted 
mean Lorentz factor $\gamma\sim 50$, we
get $\langle\zeta_{v}^{*}\rangle\sim 80(n_{c}/n_{r})\delta$ and 
$\langle\zeta_{q}^{*}\rangle\sim -0.5$ (for $n_{c}/n_{r}\ll 10^{2}$; 
see Appendix for details). 
The above value of rest-frame $\gamma$ is consistent with self-absorbed sources
having brightness temperatures in a narrow range close to $\sim 10^{11}K$ \citep{rea94}.
The actual energy distribution of the radiating
particles will not be characterized by sharp energy cut-offs but will
rather be a smooth function \citep{shi00,die00}. However, the detailed
treatment of these subtleties is beyond the scope of this paper and we 
believe that the overall complexity of the problem justifies the use of our 
approximate treatment. For the above choice of parameters, 
the main contribution to rotativity comes from
cold electrons as long as $n_{c}/n_{r}\ga 5\times 10^{-3}$ (cf. eq. [A9]). 
The required levels of LP and CP can be obtained, for example,   
when $p=1$ and the combination $(n_{c}/n_{r})\delta\sim 2.5\times 10^{-3}$,  
as this assures sufficiently low mean rotativity that it does not lead to
Faraday depolarization. Bear in mind that the admixture of cold electrons
does not have to be large to explain the data. For example, 
we get the right levels of LP and CP for $n_{c}/n_{r}\sim 5\times 10^{-2}$ 
and $\delta\sim 5\times 10^{-2}$.
Interestingly, a jet with such a plasma composition could carry 
roughly as small a kinetic power as the pure
electron-positron jet with the same emissivity, since the ratio of kinetic powers of an 
$e$-$p$ jet to a pure relativistic $e^{+}$-$e^{-}$ jet is
$\sim
18.4(\langle\gamma\rangle_{e^{+}e^{-}}/50)^{-1}(\gamma_{i,e^{+}e^{-}}/\gamma_{i,pe})$,
where we have assumed that protons are cold and 
have used $\alpha =0.5$ and where $e^{+}e^{-}$ and $pe$ refer to 
pair plasma and normal plasma, respectively.
This means that, in principle, one can also explain the mean levels of circular
and linear polarizations in 3C 279 for much lower values of $\gamma_{i}$
than the one used above
and therefore much higher rotativity, provided that the bias parameter $\delta$ is
also much lower. For example, the right levels of CP and LP can be obtained
for $\gamma_{i}=3$, $\gamma =50$, $N=1875$, $\delta\sim 3\times 10^{-3}$ 
and $\tau=1$ (see middle panels in Fig. 5).
We note that observational constraints on jet composition based on energetics
should be taken with caution in any case, as  
kinetic luminosities derived from these methods scale
with high powers of poorly determined observational quantities.

\subsubsection{Electron-positron jet}
The alternative possibility is that the jet is dominated 
by relativistic pair plasma. For example, for $\gamma_{i}=2$ and
$\gamma =50$ we get $\langle\zeta_{q}^{*}\rangle\sim -3.1$ and
$\langle\zeta_{v}^{*}\rangle\sim 1.4\times 10^{2}\delta (n_{p}/n_{e})$, where
$n_{p}$ and $n_{e}$ are the number densities of protons and electrons,
respectively. Agreement with the observed fractional 
linear and circular polarizations can be
obtained for $p=1$ and $(n_{p}/n_{e})\delta\sim 2.1\times 10^{-3}$. 
Depending on the actual values of $\delta$ and 
the ratio of protons to electrons, the jet may be pair-dominated 
in the sense that $n_{e}\gg n_{p}$ while being dominated dynamically by
protons; or it can be dominated by pairs both numberwise and dynamically.
Recent theoretical work of 
\citet{sik00} suggests that jets may be pair-dominated numberwise
but still dynamically dominated by protons.

\subsection{The case of Sgr A$^{*}$}
Among the most intriguing sets of polarization observations are 
those of the Galactic center. Observations of stellar proper motions 
in the vicinity of the nonthermal source in the center of the Galaxy 
(Sgr A$^{*}$) reveal the presence of a $\sim2.6\times 10^{6}M_{\odot}$ compact
object --- the most convincing candidate for a supermassive black hole 
\citep{eck96,eck97,ghe98}. 
Recently \citet{bow99} reported the detection of circular polarization
from Sgr A$^{*}$ with the VLA, which was confirmed by \citet{sau99}
using ATCA. The typical level of CP in their observations was $\sim 0.3\%$,
greater than the level of linear polarization. This result
may seem surprising in light of the strong limits on the ratio of CP to
LP in AGN where CP/LP is usually much less than unity \citep{wei83}. 
However, as explained above, an excess of CP over LP can be explained easily in the
framework of our model. 
Archival VLA data indicate that the mean CP was stable over ten
years \citep{bow00}. This is also not surprising as our model naturally
predicts a persistent CP sign provided that the number of field reversals,
either along one line-of-sight or across the beam area, is
sufficiently large and the source does not undergo dramatic changes from
synchrotron thin to synchrotron thick regimes. 
However, short-term circular polarization variability 
($\sim$ a few days) is also present. Interestingly, recent VLA
observations from 1.4 to 15 GHz separated by a week revealed a CP
increase at frequencies greater than 5 GHz,
which coincided with an increase in the total intensity \citep{bow00b}. 
The CP spectrum was characterized by a flat to slightly positive spectral index
($\pi_{c}\propto \nu^{\beta}, \beta\ga 0$). 
This result can also be accounted for in our model.
For example, in the framework of a self-absorbed, 
self-similar jet model \citep{bla79} but with a small bias $\delta$
we have $\langle\zeta_{v}^{*}\rangle\propto\delta\propto B\propto \nu$. 
Assuming that $\xi\ga\epsilon_{q}\zeta_{q}$ and 
$q\equiv (2p-1)^{2}\langle\zeta_{q}^{*}\rangle^{2}\gg 1$, and then demanding
$|\pi_{c}|/\pi_{l}\ga 1$ and $\beta\ga 0$, we obtain the following criterion
which leads to the observed behavior (cf. eq. [7]--[10]):\\
\begin{equation}
(\xi +q)^{2}q^{-1}\la (\delta\zeta)^{2}\la 2\xi +q+\xi (\xi +q) .
\end{equation} 
For instance, for $\xi=20$, $\langle\zeta_{q}^{*}\rangle
\sim -10$, 
$\delta\zeta\sim 40$ and $p=1$ we have $\beta\ga 0$,
$|\pi_{c}|/\pi_{l}\sim 3$ and $|\pi_{c}|\sim 0.6\%$.
Such values of the mean convertibility and rotativity could be obtained 
in the case of predominantly cold $e^{-}$-$p$ 
plasma with a small admixture of relativistic electrons. Note, however,
that a certain
amount of relativistic particles is always essential as synchrotron processes
dominate emission and absorption in typical nonthermal sources.
Specifically, the minimum and maximum size of Sgr A$^{*}$
constrain the brightness temperature to be
$10^{10}\la T_{b}\la 5\times 10^{11}K$ \citep{mel01}, which is within the
range of typical AGN radio cores.
Taking $T_{b}\sim 10^{11}$ as the representative rest frame 
value \citep{rea94}, we get $\gamma\sim 50$. 
For minimum cut-off Lorentz factor $\gamma_{i}=2$, this gives
relativistic contributions to convertibility and rotativity equal to 
$\langle\zeta_{q}^{*({r)}}\rangle\sim -3$ and
$\langle\zeta_{v}^{*({r)}}\rangle\sim 275\delta$, respectively.
Provided that dielectric suppression (Razin effect) is unimportant,
the overall transfer coefficients are merely the 
sums of cold $(c)$ and relativistic $(r)$ contributions. Thus, for 
$\zeta_{v}^{*(c)}/\zeta_{v}^{*(r)}\sim 4.3
(n_{c}/n_{r})$ and $\zeta_{q}^{*(c)}/\zeta_{q}^{*(r)}\sim
0.08(n_{c}/n_{r})$ 
(see Appendix), we obtain $(n_{c}/n_{r})\sim
30$ and $\delta\sim 10^{-3}$, which corresponds to the required
values of the mean convertibility and rotativity.\\
\indent
It has recently been suggested that observations of linear polarization 
can be used to constrain the accretion rate in Sgr A$^{*}$ and other
low-luminosity AGN \citep{ago00,qua00}. 
These authors base their argument on the assumption
that the Faraday rotation measure has to be sufficiently small in order not
to suppress strong linear polarization \citep{ait00}. This assumption 
places limits on density and magnetic field strength and leads to 
very low accretion rates $\sim 10^{-8}$ to $10^{-9}M_{\odot}$ yr$^{-1}$. 
As noted by \citet{ago00} and \citet{qua00}, 
this is inconsistent with an advection-dominated model for Sgr A$^{*}$,
which assumes that the accretion rate is of order 
the canonical Bondi rate $\sim 10^{-4}$ to $10^{-5}M_{\odot}$ yr$^{-1}$. 
We point out that strong rotation measure does not in principle 
limit densities and magnetic fields provided that the
field has a very small bias superimposed on it,
which is required to define the sign of circular polarization,
and that the dispersion of the projected magnetic field on the sky is not
too large. 
This implies that `high' accretion rates comparable to the Bondi rate cannot be excluded
on these grounds. More work on this issue is required to fully exploit the
information contained in polarization observations in order to
constrain physical conditions in Sgr A$^{*}$.

\subsection{Radio galaxy M81$^{*}$}
\citet{bru01} detected circular polarization in the compact radio jet of
the nearby spiral galaxy M81. Their estimated values of CP were 
$0.27\pm 0.06\pm 0.07\%$ at 4.8 GHz and $0.54\pm 0.06\pm 0.07\%$ at 8.4
GHz, where
errors are separated into statistical and systematic terms. This suggests
that the CP spectrum is flat or possibly inverted. They also detected no
linear polarization at a level of $0.1\%$, indicating that the source has a
high circular-to-linear polarization ratio. The spectral index indicates
that this source is synchrotron thick and we can apply the same approach 
as for Sgr A$^{*}$. For example, for $\xi\sim 50$ and $\delta\zeta\sim 80$
we get a flat to slightly positive CP spectral index, CP/LP ratio $\sim 4.7$
and $\pi_{c}\sim 0.33\%$.

\subsection{IDV blazar PKS 1519-273}
\citet{mar00} reported strongly variable circular and linear polarization
from the intraday variable blazar PKS 1519-273. The source was characterized
by an inverted intensity spectrum.
The spectrum of circular polarization was roughly consistent with being
flat or inverted (see their Fig. 3c) and typical values of linear and
circular polarizations were a few percent and $\sim 1\%$, respectively.
These data can be explained, for example, in the limit of high synchrotron depth.
In order to have a roughly flat/inverted CP spectrum, we need
the bias to correspond approximately to $\delta$ near or below 
the extremum in circular polarization according to curves in Fig. 3.
Levels of linear and circular polarization, in qualitative
agreement with the observed ones, can be obtained for various combinations of $\xi$
and the dispersion of the projected random magnetic field on the sky ($2p-1$)
(cf. equations [6]--[8] and see Figures 2 and 3). Alternatively, the data
can be explained for synchrotron depth $\tau\sim 1$. In this case, 
the flatness of the synchrotron spectrum
could be explained by means of a Blandford-K\"{o}nigl
model in which the photospheric surface of constant, model-dependent,
synchrotron depth moves outward along the jet with decreasing frequency.
The required levels of CP and LP can then be obtained easily 
for the parameters considered in Fig. 5 (see middle and right panel).

\subsection{X-ray binary SS 433}
\citet{fen00} detected circular polarization from the radio jet in the famous
X-ray binary SS 433. The flux density 
spectra of circular polarization and Stokes $I$ were roughly of the form
$V\propto I\propto\nu^{-1}$. However, multiple components in the source and a lack of high
spatial resolution prevented determination of the origin of circular
polarization and the spectrum of fractional polarization.
They argued that the CP emission is likely to be produced in the innermost
regions of the binary, and that the fractional CP of this region can be as high as
$10\%$.
In this case the ratio of the fractional circular polarization to the overall
linear polarization ($\la 1\%$) would exceed unity. 
If circular polarization originates from regions characterized by
synchrotron depth $\tau\gg 1$, then the underlying synchrotron spectrum
from this region would be approximately flat/inverted \citep{par99}. Then,
the fractional CP spectrum would scale roughly $\propto\nu^{-1}$, which 
corresponds to a bias $\delta$ (and mean rotativity) 
greater than that associated with
the CP extremum. The highest circular polarization 
would then be about $3\%$. Note that in such a situation we would also expect an
excess of CP over LP even in the innermost regions of SS 433.
If CP originates from the inner parts of the binary but from 
regions of $\tau\la 1$, then the flatness of the synchrotron spectrum
could be explained by means of a Blandford-K\"{o}nigl
model.
In such a scenario, the maximum fractional CP from the central regions 
could be as high as $\sim 20\%$ (see Fig 4) and the CP spectrum could have
a negative slope $\beta$ ($\pi_{c}\propto\nu^{\beta}$; cf. eq. [19] for
$\tau_{\xi}\ll 1$).

\section{Conclusions}
We have considered the transfer of polarized synchrotron radiation in stochastic
sources by means of an analytic approach and a set of numerical
simulations, 
and have argued that Faraday conversion is the primary mechanism responsible
for the circular polarization properties of compact radio sources.
A crucial ingredient of our model is a small bias in the highly turbulent
magnetic field which accounts for the persistence of the sign of circular
polarization. This bias is direct evidence for the net magnetic flux carried by
magnetically accelerated jets (e.g., \citet{blap82,lic92}).\\
\indent
Extremely large rates of Faraday rotation, i.e., Faraday
rotation per unit synchrotron absorption depth, do not necessarily lead to
depolarization provided that the mean rate of Faraday rotation across the
source is relatively small, or in other words, that the turbulent magnetic field
possesses a very small directional bias. Indeed, a large Faraday
rotativity is required in order to explain the high ratio of circular to
linear polarization observed in some sources.
Constraints on jet composition or accretion rate, based
on the requirement that the source does not become Faraday depolarized,
may be circumvented under these conditions.\\
\indent
Gradients in Faraday rotation across turbulent
cells can lead to correlations between rotativity and Stokes $Q$ and $U$
parameters, which can result in ``correlation depolarization''.
Observed polarization levels require that the field have many reversals along
the line of sight to avoid this effect.
Statistical fluctuations of circular and linear polarizations are then
likely to be dominated by changes in the mean parameters describing the
plasma rather than by the stochastic behavior of the turbulent medium.
Variations in the mean parameters are unlikely to change the helicity of
circular polarization unless a source undergoes a sharp transition from
very low to very high synchrotron depth.\\
\indent
We have shown that our model is potentially applicable to a wide range of
compact synchrotron sources. In particular, it naturally predicts an
excess of circular over linear polarization when a source is strongly
depolarized by the mean Faraday rotation and when a small amount of linear
polarizarion is efficiently converted into circular polarization.
This can explain the polarization properties of the Galactic Center and
M81$^{*}$. 

\acknowledgments
We thank Roger Blandford, Avery Broderick and Marek Sikora for insightful discussions. 
This work was supported in part by NSF grant AST-9876887.

\newpage
\appendix
\section{Transfer of polarized radiation}
The transfer equation of polarized radiation reads \citep{zhe74,jon88}
\footnote{Our formulae are free from typographic errors found in \citet{jon88}.}:

\begin{equation}
\left(\begin{array}{cccc}
\left(\frac{d}{d\tau}+1\right) & \zeta_{q}\cos 2\phi  & -\zeta_{q}\sin 2\phi &\zeta_{v}\\
\zeta_{q}\cos 2\phi  & \left(\frac{d}{d\tau}+1\right) &  \zeta_{v}^{*}    &\zeta_{q}^{*}\sin 2\phi\\
-\zeta_{q}\sin 2\phi & -\zeta_{v}^{*}&\left(\frac{d}{d\tau}+1\right)      &\zeta_{q}^{*}\cos 2\phi\\
\zeta_{v}            & -\zeta_{q}^{*}\sin 2\phi & -\zeta_{q}^{*}\cos 2\phi&\left(\frac{d}{d\tau}+1\right)    
\end{array}\right)
\left(
\begin{array}{c}
I\\
Q\\
U\\
V
\end{array}
\right)
=
\left(
\begin{array}{c}
1\\
 \epsilon_{q}\cos 2\phi\\
-\epsilon_{q}\sin 2\phi\\
 \epsilon_{v}
\end{array}
\right)
J
\end{equation}
where $I$, $Q$, $U$ and  $V$ are the usual Stokes parameters, 
$\tau\propto (\nu_{B}/\nu)^{\alpha +5/2}\nu_{B}^{-1}$ is the
synchrotron optical depth, $J$ is the source function, $\phi$ is the
azimuthal projection angle of the magnetic field on the sky and the coefficients of
emissivity ($\epsilon_{q}$, $\epsilon_{v}$), absorptivity ($\zeta_{q}$, $\zeta_{v}$), 
convertibility ($\zeta_{q}^{*}$) and rotativity ($\zeta_{v}^{*}$) are given by:

\begin{eqnarray}
\epsilon_{q} & = & \epsilon_{\alpha}^{q}\\
\epsilon_{v} & = & 0\\
\zeta_{q} & = & \zeta_{\alpha}^{q}\\
\zeta_{v} & = & 0\\
\zeta_{q}^{*} & = &
-\left(1+\frac{\zeta_{q}^{*(c)}}{\zeta_{q}^{*(r)}}\right)
\zeta_{\alpha}^{*q}\left(\frac{\nu_{B}}{\nu}\right)^{\alpha
-1/2}\left[1-\left(\frac{\nu_{i}}{\nu}\right)^{\alpha
-1/2}\right]\left(\alpha -\frac{1}{2}\right)^{-1},\hspace{1cm}\alpha>\frac{1}{2}\\
\zeta_{v}^{*} & = &\left(1+\frac{\zeta_{v}^{*(c)}}{\zeta_{v}^{*(r)}}\right)
\zeta_{\alpha}^{*v}\left(\frac{\nu}{\nu_{i}}\right)^{\alpha
+1/2}\frac{\ln\gamma_{i}}{\gamma_{i}}\left(\frac{n_{r}^{-}-n_{r}^{+}}
{n_{r}^{-}+n_{r}^{+}}\right)\cot\theta
\end{eqnarray}
where $\zeta_{v,q}^{*(c)}/\zeta_{v,q}^{*(r)}$ are the ratios of
convertibilities $(q)$ and rotativities $(v)$ of the cold $(c)$ and relativistic $(r)$
plasmas given by:
\begin{eqnarray}
\frac{\zeta_{q}^{*(c)}}{\zeta_{q}^{*(r)}}&=&\frac{1}{2\alpha}\;\frac{\alpha
-1/2}{1-(\nu/\nu_{i})^{\alpha -1/2}}\;\frac{1}{\gamma_{i}}\;\frac{n_{c}^{-}+n_{c}^{+}}{n_{r}^{-}+n_{r}^{+}}\\
\frac{\zeta_{v}^{*(c)}}{\zeta_{v}^{*(r)}}&=&\frac{1}{2\alpha}\;\frac{\alpha+ 1}{\alpha
+3/2}\;\frac{\gamma_{i}^{2}}{\ln\gamma_{i}}\;\frac{n_{c}^{-}-n_{c}^{+}}{n_{r}^{-}-n_{r}^{+}}.
\end{eqnarray}
\noindent
The transfer coefficients have been generalized to include
contributions from both electron and positron plasmas but they assume an isotropic
pitch-angle distribution of the radiating particles. We also neglected
circular emissivity $\epsilon_{v}$ and absorptivity $\zeta_{v}$ \citep{jon77a}.
In the above equations, $n_{c,r}^{+,-}$ are number densities of
cold/relativistic electrons $(-)$ or positrons $(+)$, 
$\nu_{B}=eB\sin\theta/2\pi m_{e}c$ is the Larmor
frequency, $\gamma_{i}$ is the low-energy cut-off 
Lorentz factor of relativistic
particles, $\nu_{i}=\gamma_{i}^{2}\nu_{B}$ is the frequency
corresponding to radiating particles of energy $\sim\gamma_{i}m_{e}c^{2}$,
$\alpha$ is the synchrotron-thin spectral index which also defines the slope of 
the relativistic particle energy
distribution $n(\gamma)\propto\gamma^{-2\alpha -1}$, 
$\theta$ is the angle between the line of sight and the direction of the 
magnetic field and  
$\epsilon_{\alpha}^{q}$, 
$\zeta_{\alpha}^{q}$,
$\zeta_{\alpha}^{*q}$ and $\zeta_{\alpha}^{*v}$ 
are the proportionality coefficients which are tabulated in \citet{jon77a}.\\
\indent
We integrated the radiative transfer equations using the Cash-Karp embedded
Runge-Kutta fifth-order method with an 
adaptive stepsize control \citep{pre92}. 
Computations were performed by integrating transfer equations in a
 piece-wise homogeneous medium. The physical
conditions in each cell, i.e., orientation of the magnetic field vector, 
were chosen using the random number generation method of 
L'Ecuyer in the implementation provided by \citet{pre92}.
This method is particularly suited for our purposes as it generates long-period 
sequences of random numbers and thus prevents any spurious correlations.
We assumed a constant strength of the variable $\mathbf{B}$-field component
but allowed its solid angle distribution to be uniform within $\theta\in
(0^{\rm o}, 180^{\rm o})$ and $\phi\in (-\phi_{o},\phi_{o})$, where
$\langle\cos2\phi\rangle_{(-\phi_{o},\phi_{o})}=2p-1$ is the average of
$\cos 2\phi$ in the interval $(-\phi_{o},\phi_{o})$. We also superimposed a weak
constant and unidirectional component $|\mathbf{B}_{u}|=\delta |\mathbf{B}|$
on the variable magnetic field.

\clearpage
\onecolumn
\begin{figure}
\plotone{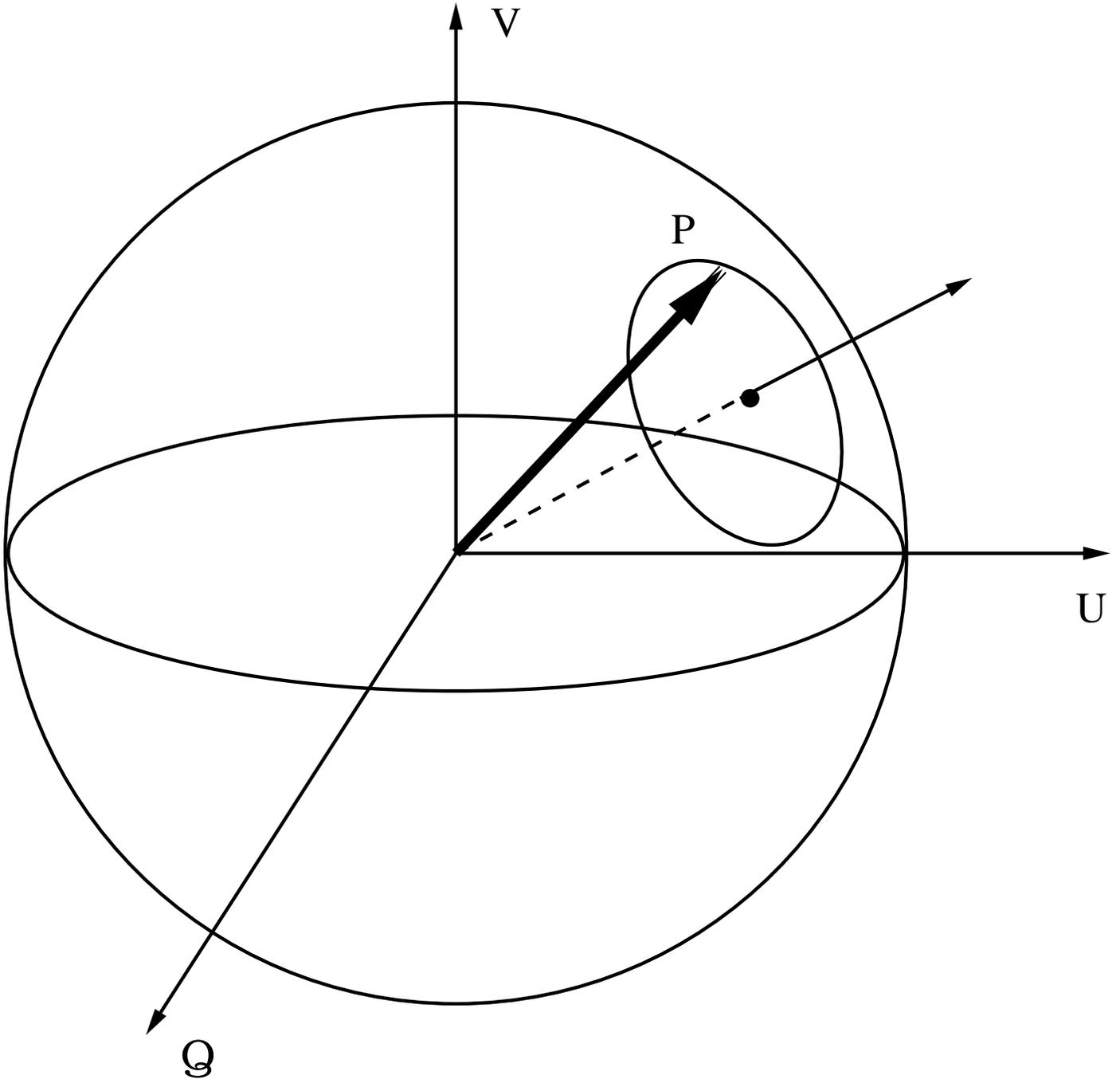}
\caption{The Poincar\'{e} sphere with the natural mode axis (dashed line)
and the polarization vector $\mathbf P$ (thick solid line). Circular and
linear 
polarizations correspond to the poles and the equator, respectively.
\label{fig1}}
\end{figure}

\begin{figure}
\plotone{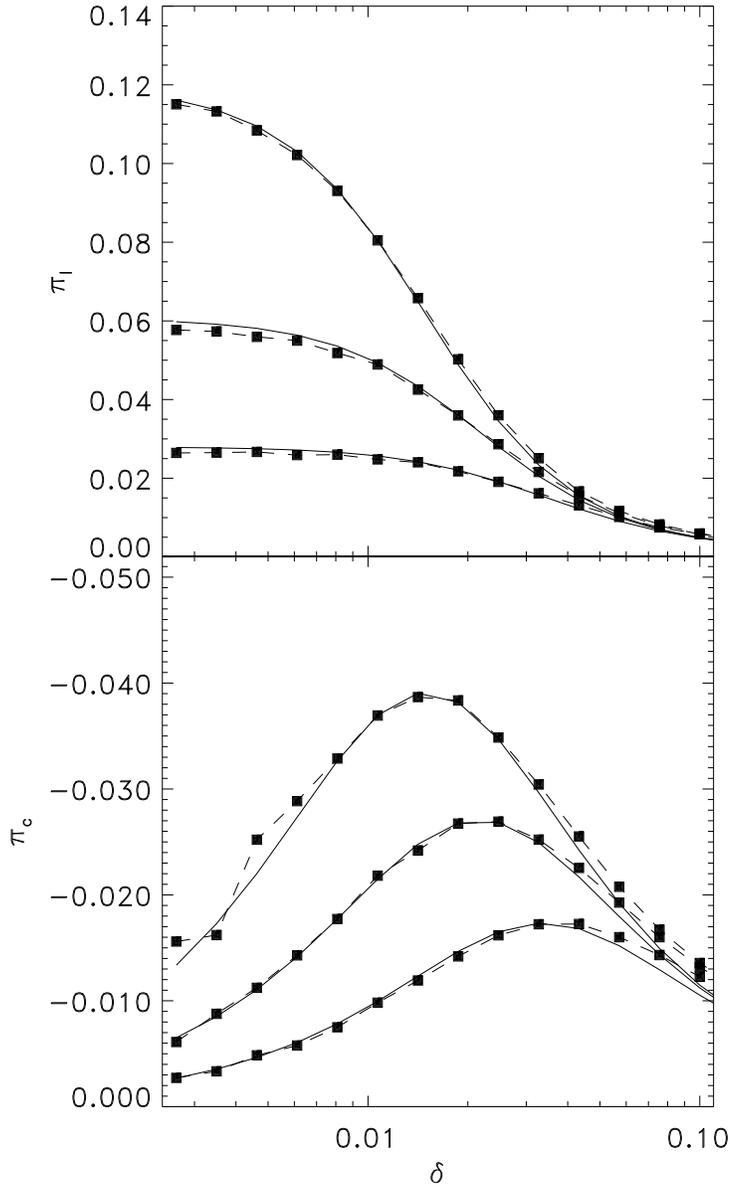}
\caption{Linear (upper panel) and circular polarizations for high
synchrotron depth ($\tau_{\rm o}\approx 27$) and varying number of
turbulent zones along the line of sight
($N=10^{8}$, $3\times 10^{5}$ and $10^{5}$; from top to bottom,
respectively). Other parameters used in this simulation were:
$\alpha=0.5$, $\gamma_{i}=3$, $\gamma =50$, $p=1$ and $\theta_{u}=45^{\rm{o}}$.
Solid lines denote analytical results and squares
connected by dashed lines denote results of Monte Carlo simulations.
\label{fig2}}
\end{figure}

\begin{figure}
\plotone{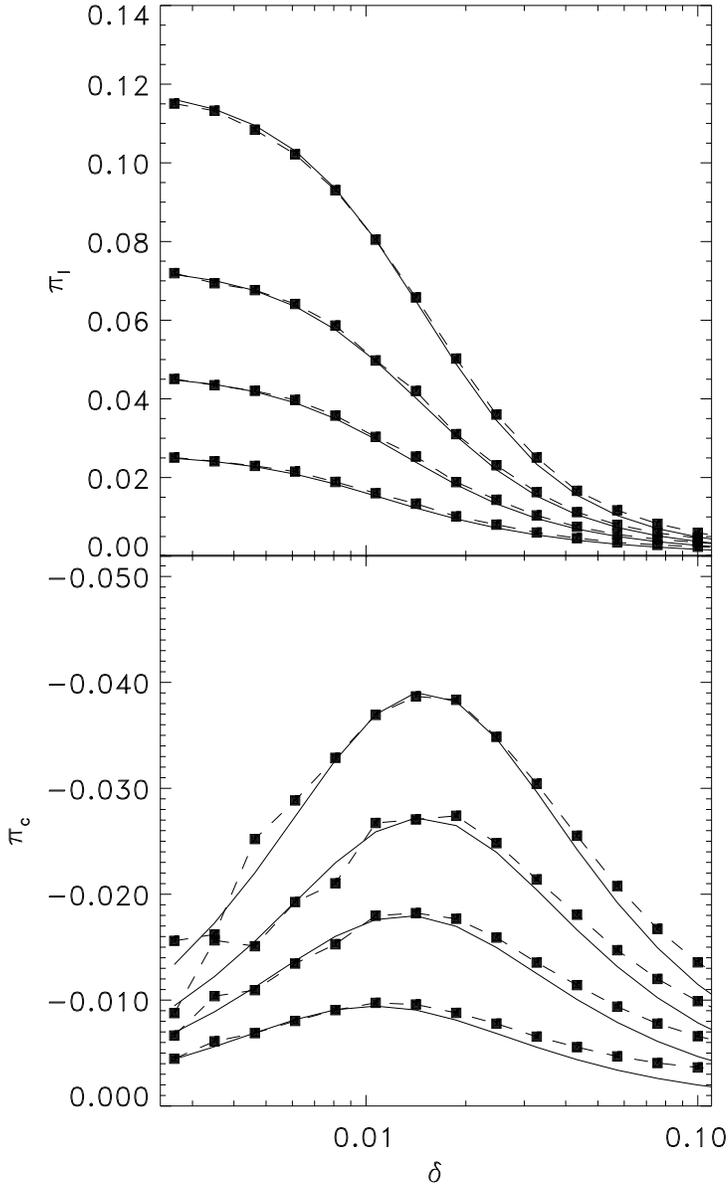}
\caption{
Linear (upper panel) and circular polarizations for high
synchrotron depth, large number of
field reversals along the line of sight (`saturated' polarization)
and varying dispersion of the projected turbulent magnetic field on the
plane of the sky ($\phi$ is distributed uniformly in $[-\Delta\phi,\Delta\phi]$;
 $\Delta\phi=0^{\rm o}$ $(p=1)$, $\Delta\phi=30^{\rm o}$ $(p=0.91)$, 
$\Delta\phi=45^{\rm o}$ $(p=0.82)$,
$\Delta\phi=60^{\rm o}$ $(p=0.71)$; from top to bottom, respectively. 
Other parameters used in this simulation were:
$\alpha=0.5$, $\gamma_{i}=3$, $\gamma =50$ and $\theta_{u}=45^{\rm{o}}$.
Solid lines denote analytical results and squares
connected by dashed lines denote results of Monte Carlo simulations.
\label{fig3}}
\end{figure}

\begin{figure}
\plotone{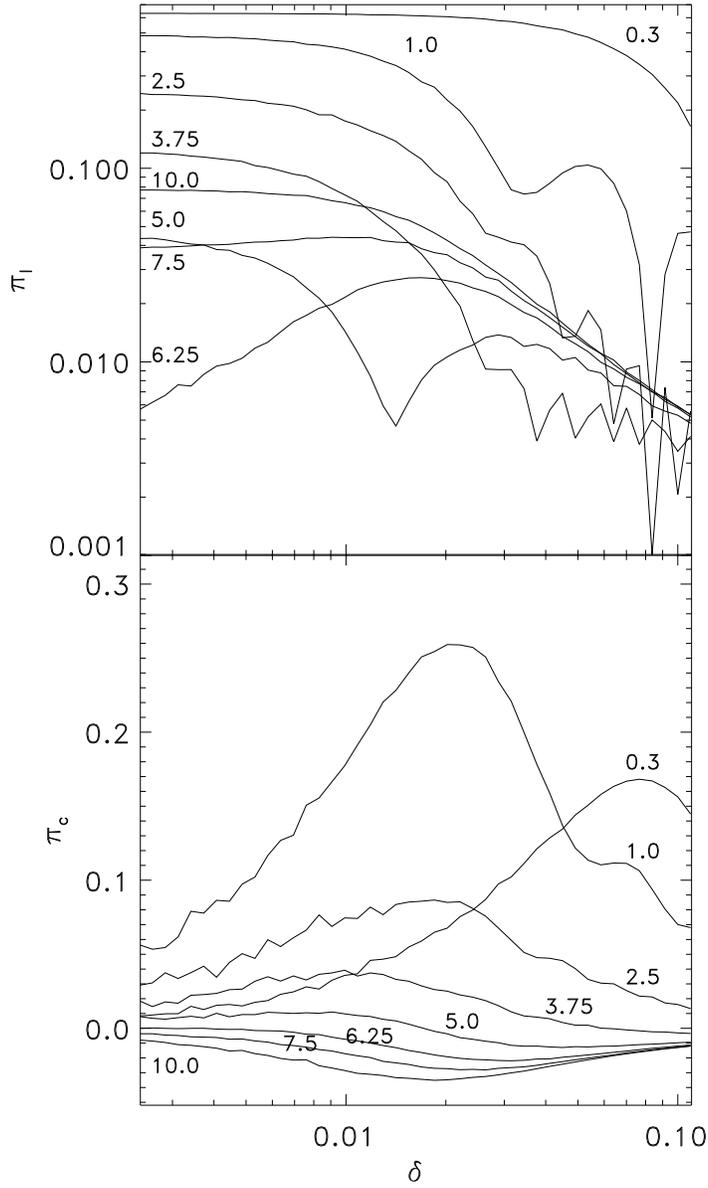}
\caption{Linear (upper panel) and circular polarizations (lower panel) for
a large number of
field reversals along the line of sight (`saturated' polarization) and
varying maximum synchrotron depth. Each curve is labeled with the value of the
synchrotron depth. Other parameters used in this simulation were:
$\alpha=0.5$, $\gamma_{i}=3$, $\gamma =50$, $p=1$ and $\theta_{u}=45^{\rm{o}}$.
\label{fig4}}
\end{figure}

\begin{figure}
\plotone{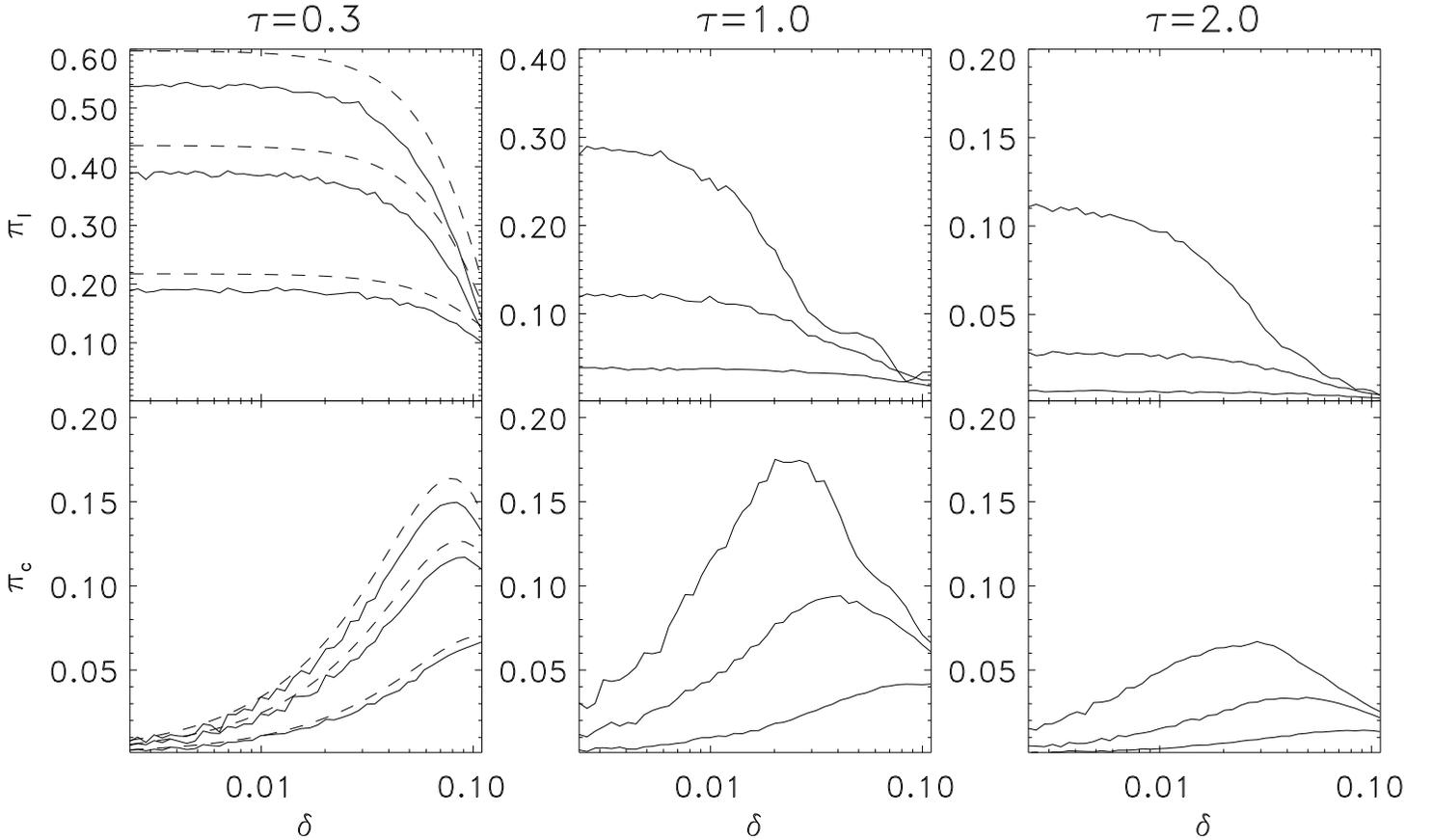}
\caption{Linear (upper panels) and circular polarizations (lower panels) for moderate
values of synchrotron depth (top labels) 
and different numbers of turbulent cells $N$ along
the line of sight. The curves correspond to 
$N=187$, $N=562$ and $N=1875$ (left panel; from bottom to top, respectively),
$N=625$, $N=1875$ and $N=6250$ (middle panel; from bottom to top,
respectively) and 
$N=1250$, $N=3750$ and $N=12500$ (right panel; from bottom to top,
respectively). 
Other parameters used in this simulation were:
$\alpha=0.5$, $\gamma_{i}=3$, $\gamma =50$, $p=1$ and $\theta_{u}=45^{\rm{o}}$.
Dashed lines on the left panel are the analytical results
for small synchrotron depth.\label{fig5}}
\end{figure}

\end{document}